\documentclass[prb,aps,epsfig,twocolumn,showpacs]{revtex4}

\usepackage{epsfig,amssymb,amsmath,latexsym}
\usepackage{subfigure}
\usepackage{wrapfig}
\usepackage{float}
\usepackage{color}
\newcommand\mbb{\mathbb}

\newcommand\beq{\begin{equation}}
\newcommand\eeq{\end{equation}}

\newcommand\bea{\begin{eqnarray}}
\newcommand\eea{\end{eqnarray}}

\newcommand\bi{\begin{itemize}}
\newcommand\ei{\end{itemize}}

\newcommand\non{\nonumber}

\newcommand\ie{{\it{i.e.}}}
\newcommand\etal{{\it{et al.}}}
\newcommand\smat{$\mathbb S$}

\newcommand\ctd{{\textsf{CT}}}

\newcommand\ct{{\textsf{CT~}}}

\newcommand\dbd{{\textsf{DB}}}

\newcommand\db{{\textsf{DB~}}}

\newcommand\sdbd{{\textsf{SDB}}}

\newcommand\sdb{{\textsf{SDB~}}}

\newcommand\nsd{{\textsf{NS}}}

\newcommand\nsnsn{{\textsf{NSNSN~}}}

\newcommand\nsnd{{\textsf{NSN}}}

\newcommand\card{{\textsf{CAR}}}
\newcommand\ard{{\textsf{AR}}}

\newcommand\ns{{\textsf{NS~}}}

\newcommand\nsn{{\textsf{NSN~}}}

\newcommand\car{{\textsf{CAR~}}}
\newcommand\ar{{\textsf{AR~}}}

\newcommand\sar{{\textsf{SAR~}}}
\newcommand\sard{{\textsf{SAR}}}

\newcommand\scar{{\textsf{SCAR~}}}
\newcommand\scard{{\textsf{SCAR}}}

\newcommand\dbdg{{\textsf{DBDG~}}}
\newcommand\dbdgd{{\textsf{DBDG}}}

\newcommand\twod{{\textsf{2D~}}}

\newif\ifboo \boofalse

\begin{document}

\textheight=23.8cm

\title{Resonant tunneling through superconducting double barrier structures in graphene}

\author{Arijit Kundu$^{1,2}$, Sumathi Rao$^{1,3}$ and Arijit Saha$^{4}$}
\affiliation{
\mbox{$^1$ Harish-Chandra Research Institute, Chhatnag Road, Jhusi, Allahabad 211019, India}\\
\mbox{$^2$ Institut f\"{u}r Theoretische Physik, Heinrich-Heine-Universit\"{a}t, D-40225 D\"{u}sseldorf, Germany}\\
\mbox{$^3$ LPTHE, Universit\'e Pierre et Marie Curie - Paris VI, 4, Place Jussieu, 75252 Paris Cedex 05, France}\\
\mbox{$^4$ Department of Condensed Matter Physics, Weizmann Institute of Science, Rehovot 76100, Israel}\\
}


\begin{abstract}
We study resonant tunneling through a superconducting double barrier structure in graphene 
as a function of the system parameters. At each barrier, due to the proximity effect, 
an incident electron can either reflect as an electron or a hole (specular as well as
retro Andreev reflection in graphene). Similarly, transport across the barriers can 
occur via electrons as well as via the crossed (specular and/or retro) Andreev channel, 
where a hole is transmitted nonlocally to the other lead. In this geometry, in the subgap 
regime, we find resonant suppression of Andreev reflection at certain energies, due to the 
formation of Andreev bound levels between the two superconducting barriers, where the 
transmission probability $T$ for electrons incident on the double barrier structure becomes unity. 
The evolution of the transport through the superconducting double barrier geometry as a function 
of the incident energy for various angles of incidence shows the damping of the resonance as 
normal reflection between the barriers increases.

\end{abstract}

\maketitle

\section{\label{sec:one} Introduction}
The discovery of graphene, a two dimensional single layer of graphite, 
by K. S. Novoselov \etal~\cite{novoselovetal1} a few years ago, has 
led to an upsurge in the study of its  transport properties, both theoretically and 
experimentally~\cite{geimreview,castronetoreview,sdsharmareview}. In graphene, there are 
six discrete points at the edges of the hexagonal Brillouin zone where the energy bands 
touch the Fermi energy, out  of which, only two are inequivalent and are commonly 
known as the $K$ and $K^{\prime}$ valleys. The low energy quasiparticle excitations about 
the $K$ and $K^{\prime}$ valleys behave like massless relativistic Dirac fermions.  
The presence of such quasiparticles in graphene provide us with an experimental 
test bed for observing many well-known phenomena in relativistic quantum mechanics, 
such as the Klein paradox~\cite{katsnelson} at low energies.

The existence of Dirac-like quasiparticles in graphene has also motivated a lot of research 
work in exploring the effects due to the proximity of a superconductor. Graphene is not a 
natural superconductor by itself. However superconductivity in a graphene layer can be induced
in the presence of a superconducting electrode near it via the proximity effect~\cite{hubert}.
A direct manifestation of proximity effect is the phenomenon of Andreev reflection 
(\ard)~\cite{andreev} in which an electron like quasi-particle incident on a
normal metal$-$superconductor (\nsd) interface is reflected back as a hole along with the 
transfer of two electrons into the superconductor as a Cooper pair. Recently it has been
predicted that a graphene \ns junction, due to the presence of the Dirac-like energy spectrum of 
its quasiparticles, can exhibit specular Andreev reflection
(\sard)~\cite{beenakker1,beenakkerreview}, in addition to the usual retroreflection observed 
in conventional \ns junctions~\cite{blonder}. The presence of the \sar process in graphene leads to 
qualitatively different behaviour in the tunneling conductance~\cite{subhro,moitri1,linder1,linder2}, 
the Josephson current~\cite{beenakker2,moitri2} and the spin current~\cite{greenbaum} as compared 
to those in conventional superconducting hybrid junctions.

An even more intriguing example where the proximity effect manifests itself is the phenomenon of
crossed Andreev reflection (\card) which can only take place in a
normal metal$-$superconductor$-$normal metal (\nsnd) junction, provided the distance between the 
two normal metals is less than or equal to the phase coherence length of the superconductor.
This is a nonlocal process where an incident electron from one of the normal leads pairs up 
with an electron from the other lead to form a Cooper pair and jumps into the superconductor. 
Due to the presence of the Dirac-like energy spectrum, like \sar, graphene can also exhibit specular
crossed Andreev reflection (\scard) in a proximity induced graphene \nsn junction. The effect
of \car in graphene has been studied earlier in Refs.~\onlinecite{cayssol,cbenjamin} in the context
of detecting entangled states in graphene. However, transport properties of a superconducting
double barrier (\sdbd) geometry in graphene, \ie~graphene \nsnsn junctions, have not been studied so far, 
where resonance effects can be more important.

In general, electronic confinement in graphene is experimentally challenging due to the effect
of Klein tunneling. In the recent past, resonant tunneling has been considered in doped graphene
(single barrier) $n$-$p$-$n$ junctions~\cite{vfalko1} and in normal double barrier 
structures~\cite{peeters,baizhang} where resonance effects on the transmission have been investigated. 
But here again, the problem of resonant transmission through doped double barriers in graphene has not
been investigated. 
Motivated by this, in this article, we study resonant tunneling through a \sdb 
structure in graphene, which, because of the correspondence between \ar
and Klein tunneling ~\cite{beenakker3} would also be valid for a doped graphene
$n$-$p$-$n$-$p$-$n$ junction.

The paper is organized as follows. In Sec.~\ref{sec:two}, we describe the set-up for our 
system, where two superconducting patches have been deposited on top of a clean graphene sheet 
to form the graphene \sdb structure and discuss its theoretical modeling. 
In Sec.~\ref{sec:three}, we obtain the resonance condition analytically
(the Andreev bound state condition) by considering multiple (retro) Andreev reflections
between the barriers. Then in Sec.~\ref{sec:four}, we discuss the effect of normal 
reflection and \sar between the barriers on the resonances for angles of incidence 
other than normal incidence. In Sec.~\ref{sec:five}, we discuss the numerical results of 
our study and show how the resonances behave as a function of the various parameters in the 
theory and also how the resonances evolve as a function of the energy for various values of the 
incident angle. Finally in Sec.~\ref{sec:six}, we present our summary and conclusions.

\section{\label{sec:two} Graphene \sdb structure}
\begin{figure}
\begin{center}
\includegraphics[width=9.5cm,height=6.0cm]{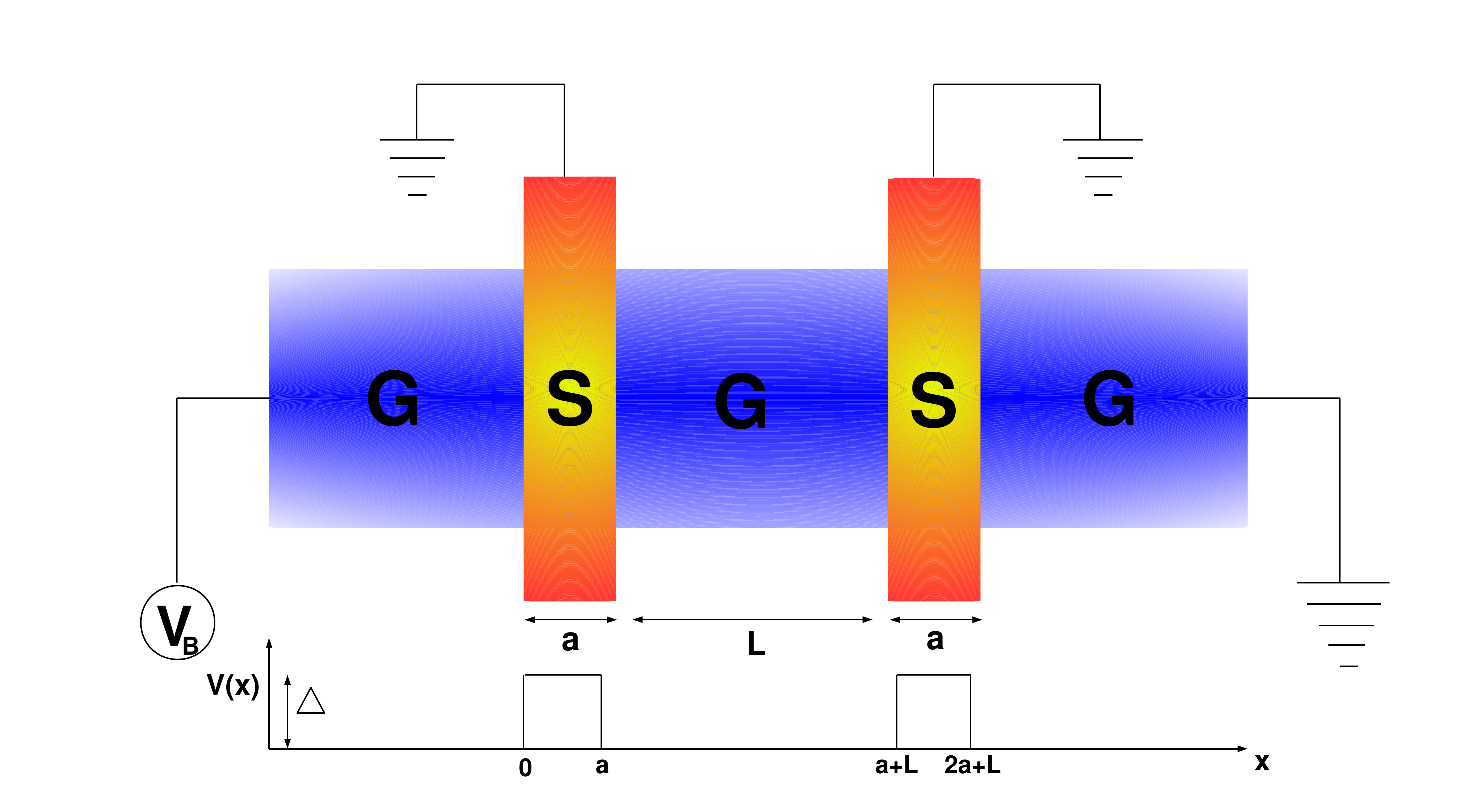}
\caption{(Color online) Cartoon of the \sdb structure in a graphene sheet. Two patches at the two places
on the graphene sheet depict superconducting material deposited on top of it.
The schematic of the potential profile seen by an incident electron is shown below.}
\label{figone}
\end{center}
\end{figure}
In our analysis, we consider a clean graphene sheet occupying the $x$-$y$ plane. 
The \sdb structure is formed by depositing thin strips of superconducting material 
on top of the graphene sheet at two places. This induces a finite superconducting 
gap ($\Delta_{i} e^{i\phi_{i}}$) in the barrier regions as a result of the 
proximity effect of the superconducting patches. Here $\Delta_{i}$ and $\phi_{i}$
are the pair potentials and order parameter phases on the two patches respectively 
($i$ refers to the index of the strips). The geometry is shown in Fig.~\ref{figone}.
The space dependence of the order parameter (which also acts as a scattering potential
for the incident electron) can be expressed as
\bea
V(x) &=& \Delta e^{i\phi} \Theta(x)
\Theta(-x+a) + \Delta e^{i\phi} \non\\
&&  \Theta[x-(a+L)] \Theta[-x+(2a+L)]
\label{potential}
\eea
where $a$ is the width of the superconducting barrier in graphene and $L$ is the distance 
between the two barriers. Here we assume that the spatial variation of 
potential steps is slow on the scale of the lattice spacing so that inter-valley scattering
is suppressed.
Also here $\Theta$ is the Heaviside $\Theta$-function, and we have taken $\phi_1=\phi_2=\phi$, 
since we will not be looking at supercurrents (Josephson effect) in this work. 

If the width $a$ of the superconducting strips is of the order of the phase coherence 
length of the superconductors, the normally incident electron can be transmitted 
across the barriers both as an electron (electron cotunneling (\ctd)) and as a hole, 
via the retro crossed Andreev reflection (\card) process which is nonlocal and occurs 
between the same bands. Importantly, besides local retro \ar and nonlocal retro \card, 
graphene can also exhibit local specular Andreev reflection (\sard) and nonlocal specular
crossed Andreev reflection (\scard)~\cite{cbenjamin} which can switch the valley bands. 
In the process of retro \card, an electron incident on the barriers is transmitted as a 
hole which retraces the trajectory of the incident electron. On the other hand, 
in \scard, a hole is transmitted nonlocally but in a specular fashion. Here, we restrict 
ourselves to spin singlet ($s$-wave) superconductors so that the electron and the hole 
are taken from  opposite spin bands in order to allow the Cooper pair to jump into the 
superconductor with net spin zero.

Because of the pseudo spin and valley degeneracies present in graphene, it suffices 
to use a four dimensional version of the Dirac-Boguliobov-de Gennes equation 
(\dbdgd)~\cite{beenakker1} for electrons and holes which is given by
\bea
\left({\begin{array}{cc} 
           \vec{k}.\vec{\sigma}-U & \Delta \\
\Delta^{\ast} & -(\vec{k}.\vec{\sigma}-U) \\
           \end{array}} \right)\left( {\begin{array}{c} 
           u \\
v \\
           \end{array}} \right)=\epsilon\left( {\begin{array}{c}
           u \\
v \\
           \end{array}} \right)
\label{dbdg}
\eea
where, $U=U({\bf r})+E_{F}$, and the energy $\epsilon$ is measured from the Fermi 
level of the superconductor. We assume that $U({\bf r})= 0$ in the normal graphene 
region and $U({\bf r}) = U_0$, a constant, independent of ${\bf r}$ in the proximity
induced superconducting region. Note that we have defined dimensionless variables 
\bea
x\Rightarrow \frac{xE_{F}}{\hbar v_{F}}, ~~~y\Rightarrow \frac{yE_{F}}{\hbar v_{F}}, ~~~k_{y} \Rightarrow \frac{\hbar v_{F}k_{y}}{E_{F}}, 
\nonumber \\
\Delta \Rightarrow \frac{\Delta}{E_{F}}, ~~~\epsilon \Rightarrow \frac{\epsilon}{E_{F}}~~{\rm  and} ~~~U \Rightarrow \frac{U}{E_{F}}
\eea
to replace the original ones. 

The solution of the \dbdg equations~\cite{beenakker1}, describing electrons and holes 
with incident energy $\epsilon$ inside the normal graphene regions ($\Delta_{(i)}=0$), 
can be written as
\bea
\Psi^{e\pm}=\frac{e^{ik_{y}y\pm ikx}}{\sqrt{\cos\alpha}}\left({\begin{array}{c}
                 e^{\mp i\alpha/2} \\
\pm e^{\pm i\alpha/2} \\ 
0 \\
0 \\
\end{array}} \right) 
\label{states1}
\eea
\bea
\Psi^{h\pm}=\frac{e^{ik_{y}y\pm ik^{\prime}x}}{\sqrt{\cos\alpha'}}\left({\begin{array}{c}
                 0 \\
0 \\ 
e^{\mp i\alpha^{\prime}/2} \\
\mp e^{\pm i\alpha^{\prime}/2} \\
\end{array}} \right)
\label{states2}
\eea
where $\alpha=\sin^{-1}[{k_{y}/(\epsilon+1)}]$, $\alpha^{\prime}=\sin^{-1}[{k_{y}/(\epsilon-1)}], 
k=\sqrt{\epsilon^{2}-k_{y}^{2}}$ and $k^{\prime}=\sqrt{\epsilon^{2}-k_{y}^{2}}$. $\alpha$ is the 
angle of incidence of the incoming electron (with wave-vector $(k,k_y)$) and $\alpha^{\prime}$ is 
the angle of reflection of the Andreev reflected hole (with wave-vector $(k^{\prime},k_y)$). 
For retro \ard, $\alpha^{\prime},k^{\prime}$ have opposite signs from $\alpha,k$ whereas 
for \sard, they have the same signs. The change from retro ($\epsilon <1$)
to \sar ($\epsilon>1$) occurs at $\epsilon =1$ (in our dimensionless units).

Similarly for the superconducting barrier regions, the four component spinor solutions 
$(u,v)$ contain electron wave-functions $u$ of one valley and hole wave-functions $v$ 
of the other valley. The \dbdg equation can now be solved for any arbitrary energy 
$\epsilon$ and the four solutions inside the superconducting barriers are given 
in the preprint version of Ref.~\onlinecite{beenakker1} 
\bea
\psi_{1/2} &=& e^{ik_{y}y\pm x \sqrt{k_{y}^{2}-(U+ \sqrt{(\epsilon^{2}-\Delta^{2} )})^{2}}}\left( {\begin{array}{c}
                 e^{i\beta} \\
\pm e^{i\beta\pm i\gamma_{1}} \\ 
e^{-i\phi} \\
\pm e^{-i\phi\pm i\gamma_{1}} \\
\end{array}} \right) 
\label{wavefunction1/2}
\eea
\bea
\psi_{3/4} &=& e^{ik_{y}y \pm x \sqrt{k_{y}^{2}-(U- \sqrt{(\epsilon^{2}-\Delta^{2} )})^{2}}}\left( {\begin{array}{c}
                 e^{-i\beta} \\
\pm e^{-i\beta \pm i\gamma_{2}} \\ 
e^{-i\phi} \\
\pm e^{-i\phi \pm i\gamma_{2}} \\
\end{array}} \right) 
\label{wavefunction3/4}
\eea
where the subscripts $1/2$ refers to the upper and lower signs on the RHS respectively,
and similarly for $3/4$ 
and 
\bea
\gamma_{1}&=&\sin^{-1}\left(\frac{k_{y}}{U+ \sqrt{(\epsilon^{2}-\Delta^{2}})}\right) \nonumber \\
\gamma_{2}&=&\sin^{-1}\left(\frac{k_{y}}{U- \sqrt{(\epsilon^{2}-\Delta^{2}})}\right)
\eea
and 
\bea
\beta&=&\cos^{-1}\frac{\epsilon}{\Delta} \ \ \ if \ \ \epsilon<\Delta \nonumber \\
&=&-i\cosh^{-1}\frac{\epsilon}{\Delta} \ \ \ if \ \ \epsilon>\Delta~.
\eea
Here, we have not taken the limit $U \gg \Delta,\epsilon$.  We have also
obtained the solution for both right-moving and left-moving electrons and holes.

Our aim now is to obtain the net quantum mechanical amplitudes for reflection, 
transmission, \ar (and \sard) and \car (and \scard) of an electron incident on 
the \sdb structure, after it has traversed both the barriers. A double barrier 
structure can always lead to resonances and this can affect the transmissions 
and the reflections through the system. For non-relativistic electrons, this scenario 
has been studied before~\cite{morpurgo,arijitkundu}. For relativistic electrons, 
the standard paradigm is that one cannot obtain confined carrier states for normal 
incidence~\cite{katsnelson,beenakker1} due to Klein tunneling. However, discrete 
energy levels can be found for carriers in graphene based quantum wires, 
as long as they have a non-zero component parallel to the barrier~\cite{peeters}.
Moreover, for normal incidence, discrete Andreev bound levels are also found between 
two superconductors in graphene~\cite{beenakker2,moitri2}. 
These levels can clearly lead to resonant transmissions in a \sdb structure in graphene.

We now follow the procedure set up in Ref.~\onlinecite{arijitkundu} and write a $4 \times 4$
\smat-matrix whose elements denote the various net reflection and transmission
amplitudes, except that now the electron and hole wave-functions, and consequently, the transmission
and reflection matrices are each of them also 4-component. Hence, the \smat-matrix for the 
\sdb structure in graphene for an incident electron with energy $\epsilon$ can be written as
\bea \mathbb{S}_{e} =
\begin{bmatrix} ~\mbb{R}_c & \mbb{R}_{Ac}& \mbb{T}_{c} & \mbb{T}_{Ac}~\\
~\mbb{R}_{Ac} & \mbb{R}_c & \mbb{T}_{Ac} & \mbb{T}_{c}~\\
~\mbb{T}_{c} & \mbb{T}_{Ac} & \mbb{R}_c & \mbb{R}_{Ac}~\\
~\mbb{T}_{Ac} & \mbb{T}_{c} & \mbb{R}_{Ac} & \mbb{R}_c~\\
\end{bmatrix} 
\label{gsmatsdb}
\eea
where $\mbb{R}_c$ stands for normal reflection of electrons or holes
and $\mbb{R}_{Ac}$ represents \ar (and \sard), which is the reflection of an 
electron as a hole or vice-versa, from both the barriers combined. Similarly, 
$\mbb{T}_c$ represents \ct or normal transmission amplitude of electrons or holes 
while $\mbb{T}_{Ac}$ corresponds to the non local \car (and \scard) amplitude for 
electron to hole conversion across the \sdb structure. As mentioned earlier, each 
of these amplitudes are themselves $4\times 4 $ matrices. The numerical results for 
these amplitudes can be obtained by matching the electron and hole wave-functions 
between the normal and proximity induced superconducting graphene at each of the 
four interfaces and the numerical results for the resonances are discussed 
in Sec.~\ref{sec:five}.
\section{\label{sec:three} Andreev bound levels}
Andreev bound states are formed due to multiple Andreev reflections (for non zero incidence angle, 
by multiple \textit{retro} andreev reflections only). Let us first consider the \ar
matrices which converts electrons to holes at each interface. The elements of the 
\ar matrix can be derived using 
\begin{eqnarray}
&&  \mathbb{R}_{A}\psi^{e+}=r_{A,he}\psi^{h-} \nonumber \\
&&  \mathbb{R}_{A}\psi^{e-}=r_{A,he}\psi^{h+} \nonumber \\
&&  \mathbb{R}_{A}\psi^{h+}=r_{A,eh}\psi^{e-} \nonumber \\
&&  \mathbb{R}_{A}\psi^{h-}=r_{A,eh}\psi^{e+}~.
\end{eqnarray}
Each of this conditions  gives two equations, and we have only 8 non zero elements in the matrix $\mathbb{R}_{A}$ if there is no reflection. More explicitly,  the matrix elements $(\mathbb{R}_{A})_{i,j}$ are:
\begin{eqnarray}
 (\mathbb{R}_{A})_{3,1}=r_{A,he}\frac{e^{i\alpha'/2}+e^{-i\alpha'/2}e^{i\alpha}}{e^{-i\alpha/2}+e^{i\alpha/2}e^{i\alpha}}\sqrt{\frac{\cos\alpha}{\cos\alpha'}} \nonumber \\
(\mathbb{R}_{A})_{3,2}=r_{A,he}\frac{e^{i\alpha'/2}e^{i\alpha}-e^{-i\alpha'/2}}{e^{i\alpha/2}e^{i\alpha}+e^{-i\alpha/2}}\sqrt{\frac{\cos\alpha}{\cos\alpha'}} \nonumber \\
 (\mathbb{R}_{A})_{4,1}=r_{A,he}\frac{e^{-i\alpha'/2}-e^{i\alpha'/2}e^{i\alpha}}{e^{-i\alpha/2}+e^{i\alpha/2}e^{i\alpha}}\sqrt{\frac{\cos\alpha}{\cos\alpha'}} \nonumber \\
(\mathbb{R}_{A})_{4,2}=r_{A,he}\frac{e^{-i\alpha'/2}e^{i\alpha}+e^{i\alpha'/2}}{e^{i\alpha/2}e^{i\alpha}+e^{-i\alpha/2}}\sqrt{\frac{\cos\alpha}{\cos\alpha'}}
\end{eqnarray}
The other elements of the matrix $ \mathbb{R}_{A}$ are found by interchanging $\alpha\leftrightarrow -\alpha^{\prime}$, $r_{A,he}\rightarrow r_{A,eh}$ and also changing their positions to the other off-diagonal block.
In particular, for $\alpha=0$, the matrix is much more simple and can be written as:
\begin{eqnarray}
\mathbb{R}_{A}=\left( {\begin{array}{cccc}
                 0 & 0 & r_{A,eh} & 0 \\
0 & 0 & 0 & r_{A,eh} \\ 
r_{A,he} & 0 & 0 & 0 \\
0 & r_{A,he} & 0 & 0\\
\end{array}} \right)
\end{eqnarray}
Note that $\alpha$ needs to be small in order to ensure that the reflection is mainly retro-reflection.
We will consider the effect of specular reflection (along with \sard) in the next
section. In our analysis we have used left-right symmetry, but we have been careful to maintain 
the distinction between electron and hole parameters, since exact electron-hole symmetry only 
exists at $\epsilon=0$. At any finite energy, the symmetry is broken.
\begin{figure}
\begin{center}
\includegraphics[width=9.5cm,height=7.5cm]{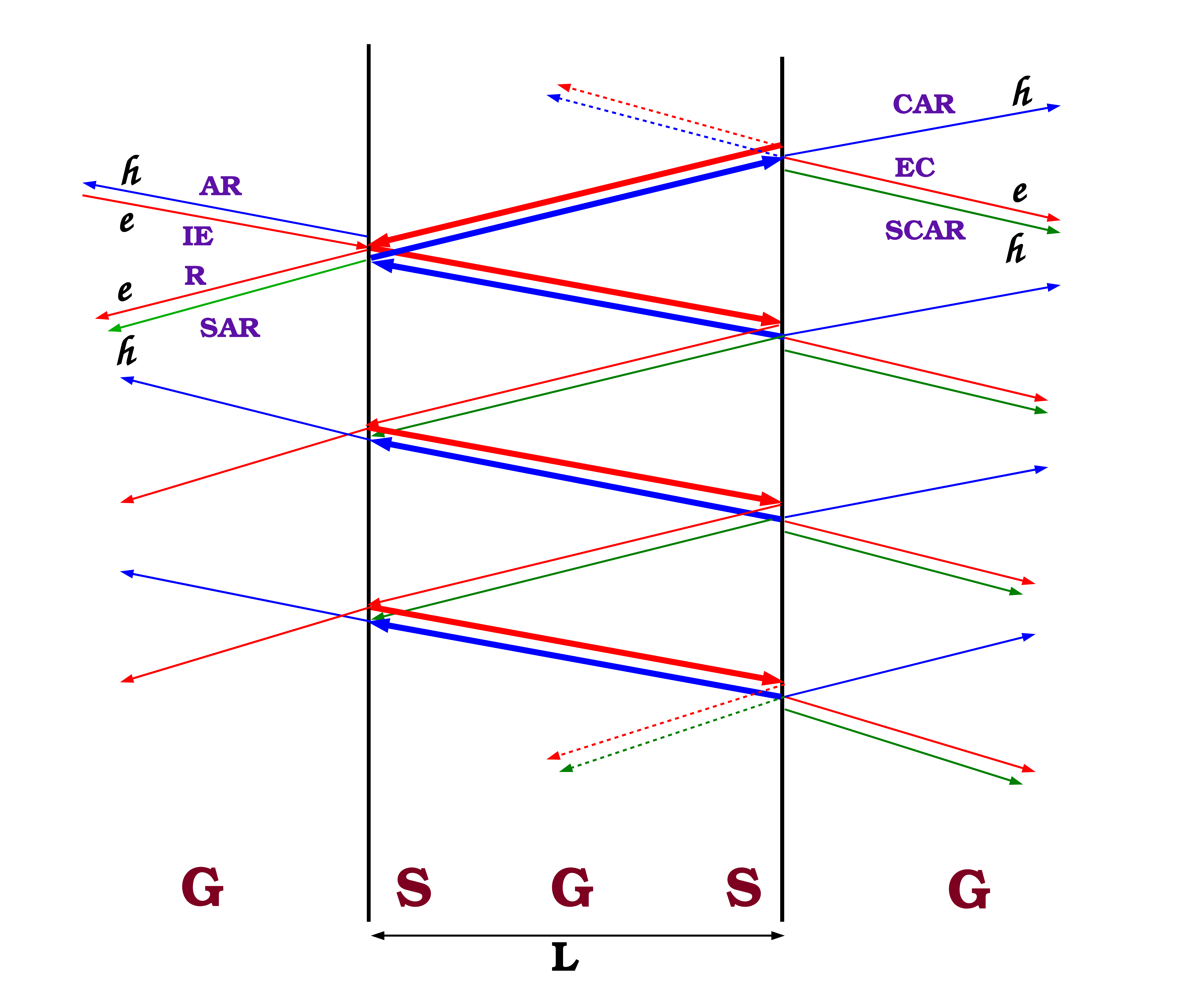}
\caption{(Color online) The electron and hole paths contributing to the formation of
Andreev bound levels between the two superconductors. The electrons have been
shown as red lines, the retro \ard, \car holes as blue lines and the 
\sard, \scar holes as green lines. The bound levels formed by multiple retro 
\ar have been shown as thick red and blue lines.  However,   not all possible 
paths have been shown in the figure.}
\label{figtwo}
\end{center}
\end{figure}
We can also write down the $4\times 4$ transmission matrix $\mathbb{T}_B$ 
which allows for both normal transmission and \car through a single barrier.  
This matrix is defined by
\begin{eqnarray}
 && \mathbb{T}_B\psi^{e\pm}=t_{e}\psi^{e\pm}+t_{A,e}\psi^{h\pm} \nonumber \\
 && \mathbb{T}_B\psi^{h\pm}=t_{h}\psi^{h\pm}+t_{A,h}\psi^{e\pm}
\end{eqnarray}~
and we find that the matrix elements are given by 
\begin{widetext}
\begin{equation}
 \mathbb{T}_B=\frac{1}{\sqrt{\cos\alpha\cos\alpha'}}\left(\begin{array}{l l l l}
t_{e} & 0 & t_{A,h}\cos(\frac{\alpha+\alpha'}{2}) & it_{A,h}\sin(\frac{\alpha-\alpha'}{2}) \\
0 & t_{e} & it_{A,h}\sin(\frac{\alpha-\alpha'}{2}) & -t_{A,h}\cos(\frac{\alpha+\alpha'}{2}) \\
t_{A,e}\cos(\frac{\alpha+\alpha'}{2}) & it_{A,e}\sin(\frac{\alpha-\alpha'}{2}) & t_{h} & 0 \\
it_{A,e}\sin(\frac{\alpha-\alpha'}{2}) & -t_{A,e}\cos(\frac{\alpha+\alpha'}{2}) & 0 & t_{h} \\
         \end{array}~
\right)
\end{equation}
\end{widetext}

Clearly $\mathbb{T}_B$ includes both normal transmission and \card.
It is also clear that $\mathbb{T}_B$ can be written as $\mathbb{T}_B=
\mathbb{T}+\mathbb{T}_A$ where $\mathbb{T}$ is a $4\times 4$ matrix with 
two non-zero $2\times 2$ diagonal blocks and $\mathbb{T}_A$ is a 
$4\times 4$ matrix with two non-zero $2\times 2$ non-diagonal blocks.

The phase matrix relating the electron and hole wave-function
when it traverses the normal graphene region through a distance
$L$ (the distance between the two superconductors) is given by
\bea
\mathbb{M} =\lambda^{-1}\mbb{D}\lambda
\eea
where 
\bea
\lambda = \begin{pmatrix}
\Lambda & 0 \\
0  & \Lambda' \end{pmatrix}
\eea
and $\mbb{D}$ is a diagonal matrix with the entries 
($e^{ikL}, e^{-ikL}, e^{ik'L}, e^{-ik'L}$) denoting the phases picked
up by the left and right moving electrons and holes respectively. The $\Lambda$ and $\Lambda'$
matrices which rotate the momentum operator to an arbitrary basis are given by
\begin{eqnarray}
\Lambda=\Lambda^{-1} = \frac{1}{\sqrt{2\cos\alpha}}\begin{pmatrix}
e^{-i\alpha/2} & e^{i\alpha/2} \\
e^{i\alpha/2} & -e^{-i\alpha/2}  \end{pmatrix}
\end{eqnarray}
with
\bea
\Lambda' &=& \frac{1}{\sqrt{2\cos\alpha'}}\begin{pmatrix}
e^{-i\alpha'/2} & -e^{i\alpha'/2} \\
e^{i\alpha'/2} & e^{-i\alpha'/2}  \end{pmatrix}
\eea

The condition for resonance or for a bound state in the normal
graphene region between the two superconductors is now just the
condition that the total transmission computed as
\begin{eqnarray}
\psi_{T} &=&  \mathbb{T}_B\left[\mathbb{M}+\mathbb{M}
\mathbb{R}_{A}\mathbb{M}\mathbb{R}_{A}\mathbb{M}+\dots \right] \mathbb{T}_B\psi^{e+} \nonumber \\
&=& \mathbb{T}_B \mathbb{M}\left[\mathbb{I} - \mathbb{R}_{A}\mathbb{M}\mathbb{R}_{A}\mathbb{M}
\right]^{-1}  \mathbb{T}_B\psi^{e+}
\label{totaltransmission}
 \end{eqnarray}
has a vanishing denominator. This is precisely the condition for the Andreev bound states formed by 
\textit{retro} \ard s \textit{without the presence of reflection}. 
From this condition, one can find the corresponding Andreev bound state energy levels.
Note that if we want the total transmission
of electrons, we just need to replace $\mathbb{T}_B$ by $\mathbb{T}$ and if we want
the total \car of holes, we need to replace
$\mathbb{T}_B$ by $\mathbb{T}_A$. Both of them show the effect of the resonances.
\section{\label{sec:four} Effect of reflection and \sard}
For normal incidence in graphene, due to pseudo-spin symmetry, normal reflection is prohibited
and we have pure Andreev bound states between the two superconductors. 
Normal bound states (formed by multiple ordinary reflections) are, in any case, not possible
in graphene even at any other incident angle, when ordinary reflection is allowed.
This is because multiple ordinary reflections are specular in nature and lead to a mode running 
along the $y$-axis, rather than a bound state. This is also true even if we have \sard, 
which leads to a specular Andreev mode running along the $y$-axis~\cite{beenakkerreview}.

\begin{figure}
\begin{center}
\includegraphics[width=7.0cm,height=4.0cm]{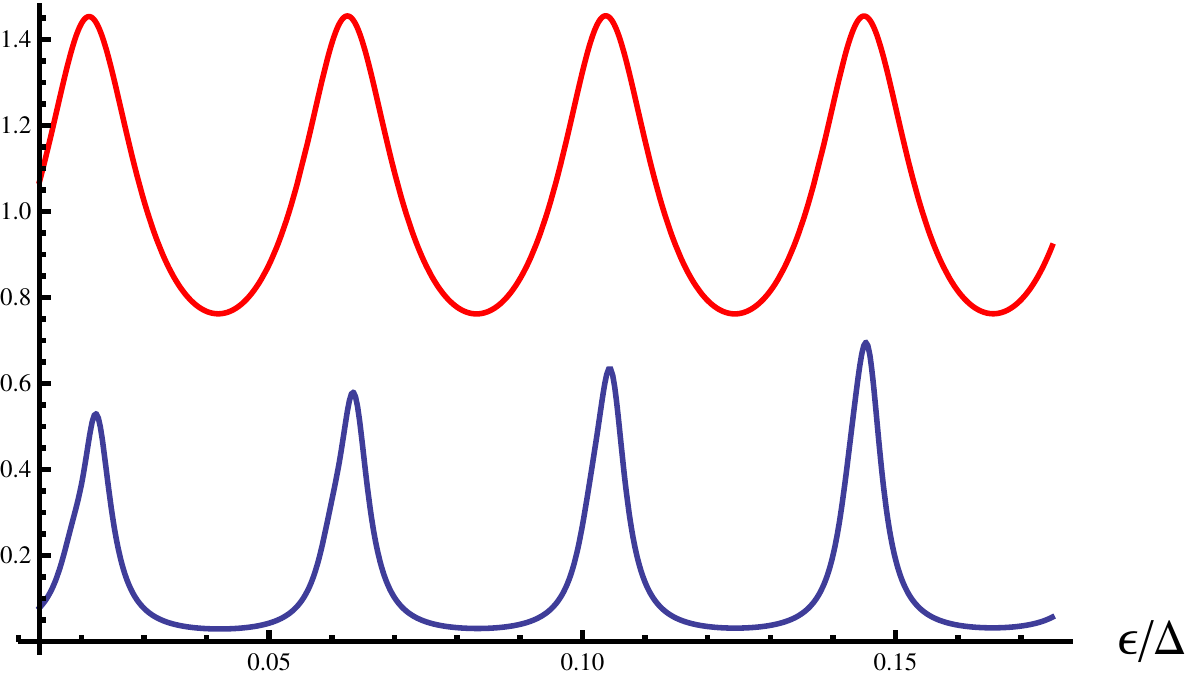}
\caption{(Color online) The bottom graph shows the behaviour of angle resolved conductance, 
obtained numerically, in units of $4e^{2}/h$ as a function of energy in the 
subgapped regime ($\epsilon\ll\Delta$) for $k_y = 0.75$. 
Here, $\Delta/E_{F}= 0.05$ and $U_{0}/E_{F} = 10.0$. 
The top graph depicts the denominator of Eq.\ref{totaltransmission} 
for the same parameter values.}
\label{figthree}
\end{center}
\end{figure}

However, at a non-zero angle of incidence, if we have specular reflection present 
(Andreev as well as normal) at each graphene \ns interface, we can still have  Andreev bound 
states formed by multiple retro \ar in between the two superconducting barriers. 
However, due to the specular nature of the reflection and the \sard, they are no longer localised. 
This is shown in Fig.~\ref{figtwo}. The incident electron can transmit through the
first superconducting barrier as an electron and then have multiple retro \ard. 
It can also have either ordinary reflection or \sar from the interface.
The transmitted electron once again (in fact, for several times, depending on the 
length of the graphene sheet in the $y$-direction) can have multiple retro \ar
before it is finally transmitted (or reflected) as an electron or a hole through the
graphene \sdb structure. The electron can also cross the first superconductor as a hole 
by retro \car and then have multiple retro-reflections from the two
interfaces and then this process can continue as well. The number of reflections 
(or \sar's) that can occur in a given sample is controlled by the length of the 
graphene sheet in the $y$-direction and the angle of incidence of the electron. 
Naively, the number should go as $L_y/L\sin\alpha$, where $L_y$ is the length of 
the sheet in the $y$-direction and $L$ and $\alpha$ have already been defined earlier. 
In Fig.~\ref{figtwo}, we have only shown some of the possible quantum 
mechanical paths to emphasize how the Andreev bound states can form between the two barriers.
We have also shown only a single incident electron, but the incident electron can also
be at any point along the $y$-axis. Hence, if we measure the total output current collected 
throughout the $y$-length of the graphene sample, we should to able to get the signature of 
the many Andreev bound states present between the two superconducting barriers. 
In fact, the signatures of the Andreev bound states should be present in all the four
amplitudes $r_c$, $r_{Ac}$, $t_c$ and $t_{Ac}$, where these four amplitudes denote
the quantum mechanical amplitudes for reflection, \ar (\sard), transmission and
\car (\scard) across the \sdb structure for an incident electron.
They should show either a maximum or a minimum
precisely at the Andreev levels obtained in Eq.~\ref{totaltransmission}. This is the main point that 
we wish to emphasize in this paper. The Andreev levels can be obtained by measuring the transmissions
and reflections through a \sdb system even for large enough $L_y$; it does not require an effective 
`one-dimensional' system. Furthermore, the signature of the resonances occurs in all the
four amplitudes. 

\begin{figure*}[ht]
  \centering
  \subfigure[$k_{y}=0$]{\includegraphics[width=0.3\textwidth]{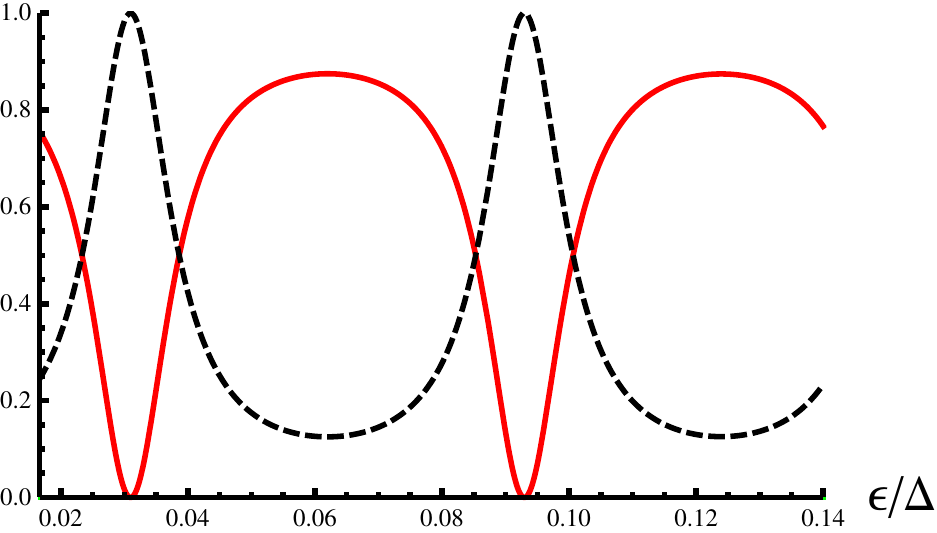}}
  \subfigure[$k_{y}=0.2$]{\includegraphics[width=0.3\textwidth]{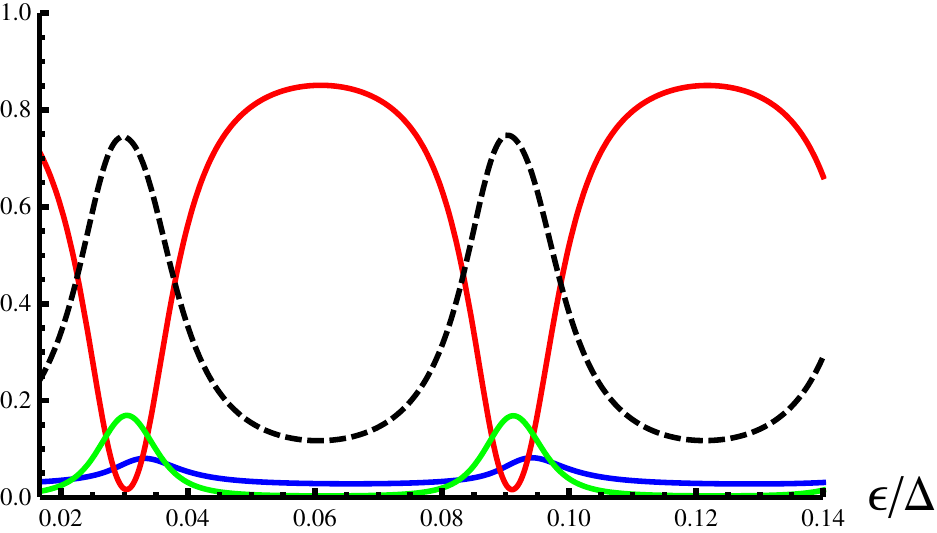}}
  \subfigure[$k_{y}=0.75$]{\includegraphics[width=0.3\textwidth]{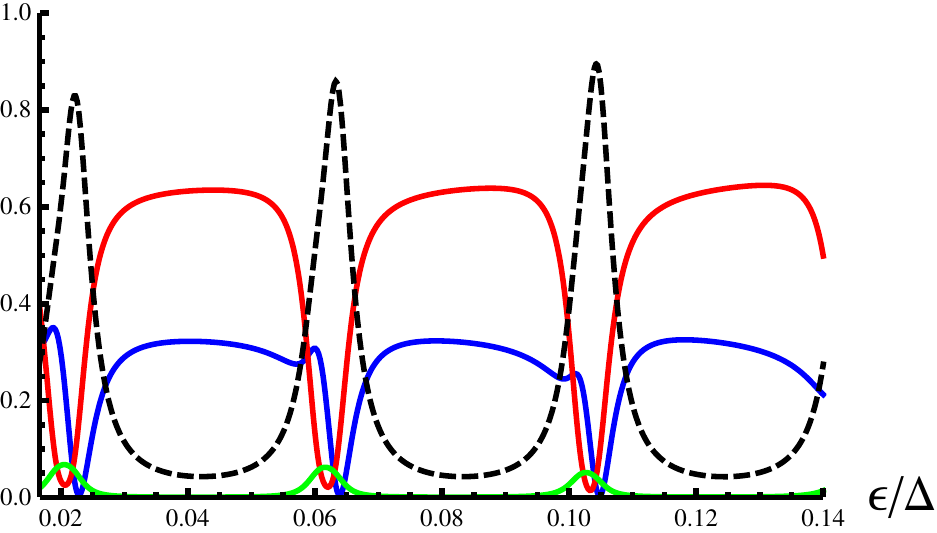}}
   \caption{(Color online) The behaviour of all possible quantum mechanical scattering probabilities 
($|r_{c}|^{2}, |t_{c}|^{2}, |r_{Ac}|^{2}, |t_{Ac}|^{2}$) through the graphene \sdb 
structure are plotted as a function of energy in the subgapped regime 
($\epsilon\ll\Delta$) for three different values of $k_y$. In (a), (b) and (c)
solid red, green, blue and dashed black lines correspond to the $|r_{Ac}|^{2}, |r_{c}|^{2}, 
|t_{Ac}|^{2}$ and $|t_{c}|^{2}$ respectively. Here, $\Delta/E_{F}= 0.05$ and 
$U_{0}/E_{F} = 10.0$.}
\label{figfour}
\end{figure*}
To illustrate the above argument, 
in Fig.\ref{figthree}, we compare the positions of the resonances as a
function of $\epsilon/\Delta$ obtained by the
vanishing of the denominator of Eq.\ref{totaltransmission}, the analytically
obtained Andreev levels, (which do not include the effects of reflection)
with the numerically obtained values of the conductance 
(by solving the scattering problem in the next section) for the same parameter values.
Note that the numerical results do include reflection since we have
chosen $k_y = 0.75$. Moreover, the numerical results include transmission
throughout all points along the $y$-axis (width of the graphene sheet). 
But as can be seen from the figure, the resonances still fall on top of each other. 
This clearly shows that the presence of specular reflection (and \sard)
has no effect on the position of the resonances which simply occur
due to the formation of the Andreev bound states discussed in the earlier
section.
\section{\label{sec:five} Numerical results}
In this section we describe the consequences of all the allowed quantum
mechanical processes across the \sdb geometry in graphene. To the left
of the \sdb structure, with an incident electron from the left, the wave-function
can be written as
\beq
\psi^{e+}+r_{c}\psi^{e-}+r_{Ac}\psi^{h-}
\eeq
and to the right of the \sdb structure, the wave-function can be written as
\beq
 t_{c}\psi^{e+}+t_{Ac}\psi^{h+}~.
\eeq
Here we use the standard wave-function matching technique
to solve such scattering problems to obtain all the four
quantum mechanical amplitudes. Hence matching the wavefunctions 
for the normal and proximity induced superconducting regions 
(Eq.(\ref{states1}-\ref{wavefunction3/4}))
at the four \ns interfaces in graphene ($x=0,a,a+L,2a+L$) forming the \sdb
structure, we obtain sixteen linear equations. Numerically solving these
sixteen equations, we obtain the four amplitudes $r_c$, $r_{Ac}$, $t_c$ and $t_{Ac}$, 
for the \sdb structure, for an incident electron with energy $\epsilon$ below the gap $\Delta$. 
Note that we distinguish between electron and hole parameters, and hence the four 
amplitudes  will be  different for incident electrons and holes.

In our numerical analysis we do not distinguish between the specular and retro  
Andreev reflections and we also allow for normal reflection at each \ns interface
in graphene, besides normal transmission and \car (specular and retro). 
The numerical results clearly show that for normal incidence of electron 
($\alpha=0$), the net normal transmission (\ctd) $t_c$ and the net \ar $r_{Ac}$ 
show resonant behaviour in the subgapped ($\epsilon\ll\Delta$) regime. 
This is shown in the first panel in Fig.~\ref{figfour} and is precisely what 
one would expect, because for normal incidence $r_{c}=0$ due to the pseudo-spin symmetry, 
and this also leads to $t_{Ac}=0$. For normally incident electrons,
this kind of resonant behaviour is completely forbidden in a normal 
double barrier (\dbd) structure in graphene due to the phenomenon of Klein tunneling.
This is the striking difference between a normal \db and a \sdb in graphene.

\begin{figure}
\begin{center}
  \subfigure[]{\includegraphics[width=0.23\textwidth]{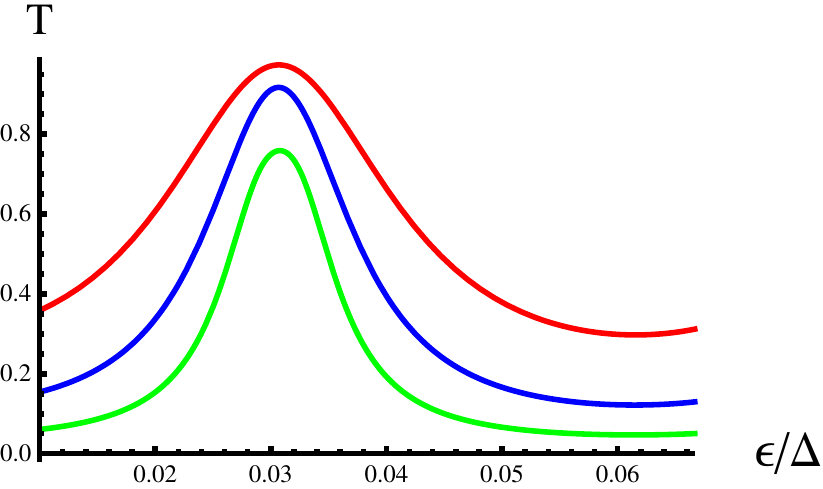}}
  \subfigure[]{\includegraphics[width=0.23\textwidth]{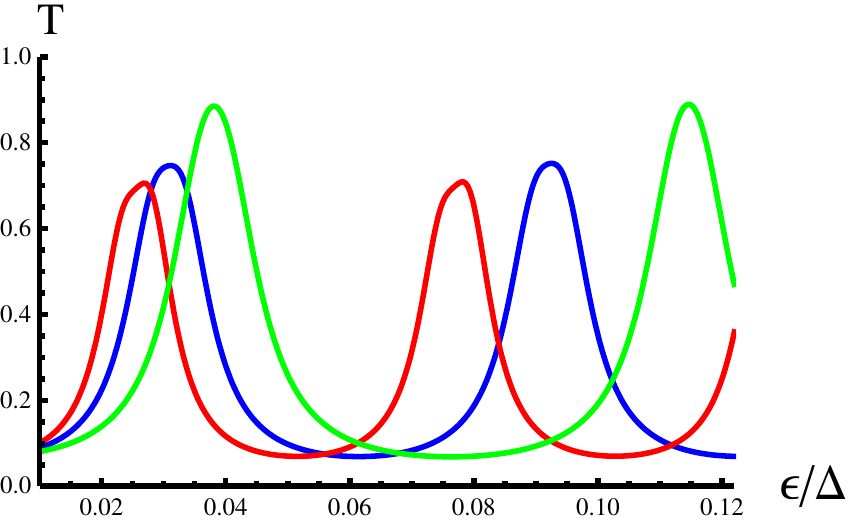}}
\caption{(Color online) (a) The behaviour of the transmission resonances ($T=|t_{c}|^{2}$) is
plotted as a function of energy in the subgapped regime ($\epsilon\ll\Delta$)
for three different values of $a/L$ ratio. The red, blue and green lines 
correspond to the three different values of $a/L$ which are 0.012, 0.017 and 0.022 
respectively. (b) The distance between consecutive resonances for the same three 
different values of $a/L$. For both the figures $k_{y}=0.125$.}
\label{figfive}
\end{center}
\end{figure}
\begin{figure*}[ht]
\centering
\subfigure[$k_{y}=0$]{
\includegraphics[width=0.3\textwidth]{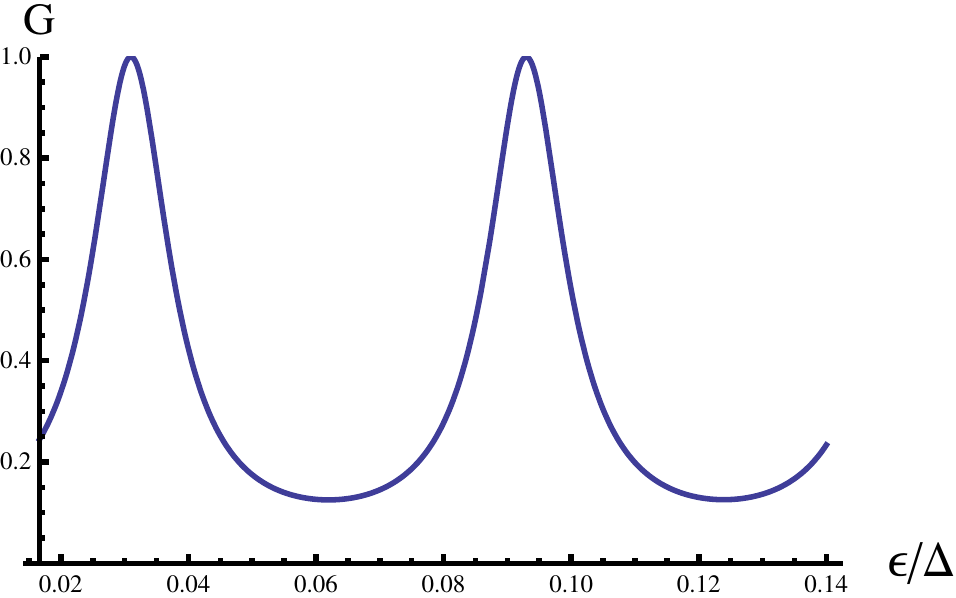}
\label{fig:subfig1}
}
\subfigure[$k_{y}=0.2$]{
\includegraphics[width=0.3\textwidth]{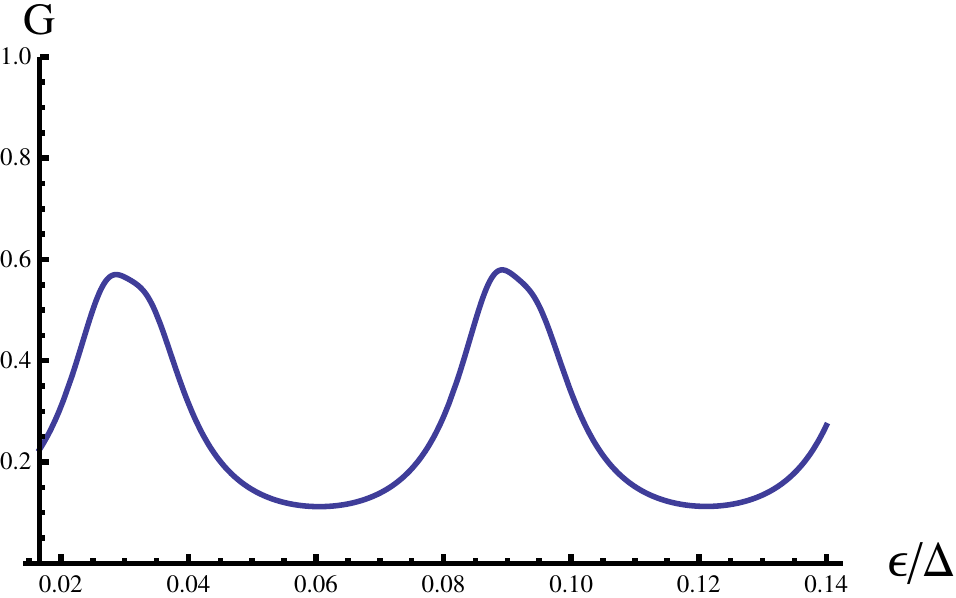}
\label{fig:subfig3}
}
\subfigure[$k_{y}=0.75$]{
\includegraphics[width=0.3\textwidth]{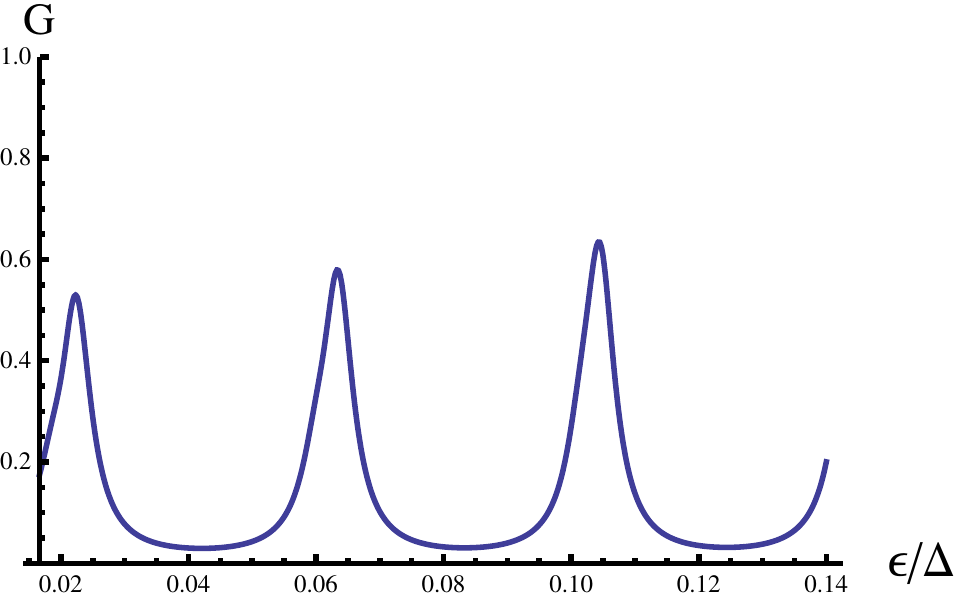}
\label{fig:subfig3}
}
\caption{(Color online) The behaviour of the angle-resolved differential conductance in units of $4e^{2}/h$ 
as a function of energy in the subgapped regime ($\epsilon\ll\Delta$) 
for three different values of $k_y$. Here, $\Delta/E_{F}= 0.05$ and 
$U_{0}/E_{F} = 10.0$.}
\label{figsix}
\end{figure*}

We also vary the momentum $k_y$ or equivalently the angle of incidence, 
and study the resonances in Fig.~\ref{figthree}. As we have already mentioned,
at $k_y=0$ or normal incidence, we have non-zero values only for $r_{Ac}$ and $t_c$, 
and strong resonant behaviour for the transmission $t_{c}$. 
As soon as the angle of incidence changes, we see the evolution 
of the $t_{c},r_{c},t_{Ac}$ and $r_{Ac}$ in the panels in Fig.~\ref{figthree}. As we increase 
the angle of incidence, due to finite $k_{y}$, normal reflection $r_{c}$ between the 
barriers increases, and due to the presence of both $r_{c}$ and $r_{Ac}$ between the 
barriers the amplitude of the transmission resonances decreases. However, at very 
large $k_y$, the roles of reflection and Andreev reflection switch; this is true 
even for a single \ns junction in graphene, since $r_{Ac}$ is proportional to 
$\cos\alpha$ and $r_{c}$ is proportional to $\sin\alpha$~\cite{beenakker1}. 
Hence, for large $k_y$, we find large values of $r_{c}$, but very small values 
of $r_{Ac}$ and once again, strong transmission resonances emerge.

These resonances can also be tuned by varying parameters such as the 
ratio of the width of the superconductor $a$ to the length between the two superconducting 
barriers $L$. As $a/L$ increases, we find that the resonances become sharper, and the 
distance between consecutive resonances increases. The behaviour of the resonances 
as a function of $a/L$ is shown in Figs~\ref{figfour}(a) and \ref{figfour}(b). 
 
The net angle-resolved differential conductance through the \sdb system is now given by
\bea
G=N*G_0\Big[|t_A|^2\cos(\alpha')-|t|^2\cos(\alpha)\Big] 
\label{DCD}
\eea
where
$N$ is the number of input channels or transverse modes in a graphene sheet of width $L_y$,
$G_0=4e^2/h$ is the unit of conductance 
and the factor of 4 comes from the pseudo-spin and valley degeneracies present in graphene.
In our numerical analysis we have considered the temperature to be zero and 
also assumed linear bias. 
We again wish to emphasise the fact that since multiple Andreev reflections 
between the two superconducting barriers not only include both specular and retro \ard, but also normal 
reflections, the exit point from the second superconductor can be anywhere along the width 
($y$-axis) of the graphene sheet. Therefore, both normal reflection as well as \sar can 
change the position of the transmitted beam along the $y$-axis at each reflection.
Hence, the total transmission here includes transmission at all points along $L_y$.
In other words, the output lead has also to be as wide as the graphene sheet.
In Fig.~\ref{figfive}, we show the net angle resolved Landauer-Buttiker conductance,
given in Eq.~\ref{DCD}, (for $N=1$) as a function of the energy of the incident electron 
$\epsilon\ll\Delta$, where again, the resonant behaviour can be seen for different 
values of $k_y$ or the incident angle. The behaviour of the  conductance,
also shows how the Andreev levels evolve as a function of the incident angle, showing 
that the height of the resonances is large when the multiple reflection between the 
barriers is either dominated by retro \ar (small angles) or normal reflection (large angles).

\section{\label{sec:six} Summary and Discussions}
To summarize, in this paper, we have computed the transmission of an electron through
a \sdb structure in graphene and shown the resonant suppression of Andreev reflection 
at certain energies below the superconducting gap $\Delta$ where normal transmission 
$|t_{c}|^{2}$ becomes unity. This resonant behaviour is absent in a normal double barrier 
in graphene due to Klein tunneling. We also show that the resonant suppression is 
due to the formation of Andreev bound states between the two superconducting barriers.
Even at finite incident angles, the position of these Andreev bound levels remain unchanged
in presence of reflection and \sard. Although, the transmission resonances get damped as the 
incident angle increases due to reflection and \sard. However, at large angles, the roles of \ar
and ordinary reflection get reversed, and once again, we see strong transmission
resonances. The point that we wish to emphasize is that at any angle
of incidence, all four transmission and reflection amplitudes, as well as the
total conductance, shows strong signals of the Andreev bound states.
We have also studied the resonances as a function of the ratios
of the  width of the superconducting barriers and the length of the 
normal graphene region between the two superconducting barriers. The most interesting 
study is the evolution of the resonances as a function of $k_y$, the momentum in the 
$y$-direction or equivalently the angle of incidence of the incoming electron.

As far as the practical realization of such a \sdb structure in graphene 
is concerned, it should be possible to fabricate such a geometry by depositing thin
strips of a spin singlet superconductor (like $Al$ or $Nb$) on top of a graphene
sheet~\cite{hubert} at two places. The width of the strips should be of the order 
of the superconducting phase coherence length ($10-15 nm$ in case of $Nb$) for \ct
and \car (and \scard) to take place. For a given $a/L$ ratio, the $|t_{c}|^{2}=1$ 
resonance in this \sdb geometry can be tuned by varying the energy of the incident 
electron (which can be done by applying a small bias voltage between the two reservoirs 
keeping within linear response, so that our calculations are valid). 
In Fig.~\ref{figfour}(a) and \ref{fig:subfig1}, the equivalent temperature at which 
the first resonance occurs is approximately $10mK$ 
(critical temperature $T_{c}\approx9.2K$ in case of $Nb$) for $a\sim10-15 nm$ and $L\sim2\mu m$.

Note also that, experimentally, it is not easy to control the angle of 
incidence of the electrons that impinge on the \sdb structure. In fact, for a given 
sample what can possibly be done is to grow the barriers at different tilted positions 
by lithographic techniques which allows the electrons to be incident on the barriers at 
different angles. However, in a given experiment, the angle of the incident electron 
is fixed, since the incident electrons are collimated. Therefore different 
experiments need to be performed with ballistic graphene samples to obtain results at 
different angles of incidence. In a given experiment, the angular spread of the incident 
electrons which would exist due to lack of perfect collimation is actually quite small, 
hence even after angular averaging our results at a given angle should qualitatively 
remain the same.

We also expect the features of the resonances to be qualititavely 
unaffected by long-ranged impurities (slow on the scale of the lattice spacing 
\ie~$k_{F}l_{m}\gg 1$ where $k_{F}\sim 1/d$, $d$ is the lattice spacing and 
$l_{m}$ is the mean-free path of electron in graphene), because such impurities 
in graphene can only cause intra-valley scattering. On the other hand, short 
range impurities ($k_{F}l_{m}\ll 1$) can cause intervalley scattering which can 
destroy the resonance. Also, strong disorder can localize electron/hole states 
between the two barriers which always can destroy the resonance. Another experimental 
variable which is expected to destroy the resonance is the presence of a magnetic field 
which bends the electron paths between the two barriers. Hence, it prevents the multiple 
retracing of the path, which is needed for resonance. Therefore, we do not expect 
resonances in the presence of strong disorder or even weak magnetic fields.

In this work, we have restricted ourselves to rectangular barriers,
whereas experimentally, the barriers are expected to be smoother~\cite{phillipkim}.
As shown in the references by Cayssol \etal~\cite{cayssol1} and Sonin~\cite{sonin},
the transmissions are affected by the shape of the potential, and it is shown that the 
resonances for normal double barriers in graphene are sharpened by smoother barriers.
By analogy, we expect our resonances to be sharpened if we make the barriers 
smooth instead of using rectangular barriers. Nevertheless, to check it explicitly, 
we need to repeat the  calculations for other shapes of barriers which is beyond
the scope of this paper.

From the application point of view, this \sdb geometry also can be used to confine
electrons in graphene which is experimentally very challenging due to Klein tunneling.
The resonant structure of $|t_{c}|^{2}$ as a function of $\epsilon/\Delta$ is
analogous to the Coulomb blockade peaks of a quantum dot. Hence, this kind of \sdb
geometry can be used to form quantum dots in graphene. Another interesting point
which can be mentioned is that this kind of proximity induced \sdb geometry can 
also be obtained by replacing the graphene sheet by a \twod Quantum Spin Hall 
topological insulator. The only difference will be the replacement of the factor 
of $4$ in the Landauer-Buttiker conductance formula by unity, since there are no 
pseudo-spin or valley degeneracies present in such quantum spin Hall insulator. 
\acknowledgments{The work of A.S. was supported by the Feinberg fellowship programme 
at WIS, Israel. One of us (A.S.) acknowledge Anindya Das for useful discussions and comments. 
One of us (S.R.) would also like to acknowledge hospitality at the Perimeter
Institute for Theoretical physics, Canada, where this work was completed.}
\bibliographystyle{apsrev} 

%
\bibliography{gsgsg_ref} 

\begin{thebibliography}{10}
\expandafter\ifx\csname bibnamefont\endcsname\relax
  \def\bibnamefont#1{#1}\fi
\expandafter\ifx\csname bibfnamefont\endcsname\relax
  \def\bibfnamefont#1{#1}\fi
\expandafter\ifx\csname url\endcsname\relax
  \def\url#1{\texttt{#1}}\fi
\expandafter\ifx\csname urlprefix\endcsname\relax\def\urlprefix{URL }\fi
\providecommand{\bibinfo}[2]{#2}
\providecommand{\eprint}[2][]{\url{#2}}

\bibitem{novoselovetal1}
\bibinfo{author}{\bibfnamefont{K.~S.} \bibnamefont{Novoselov}},
  \bibinfo{author}{\bibfnamefont{A.~K.} \bibnamefont{Geim}},
  \bibinfo{author}{\bibfnamefont{S.~V.} \bibnamefont{Morozov}},
  \bibinfo{author}{\bibfnamefont{D.}~\bibnamefont{Jiang}},
  \bibinfo{author}{\bibfnamefont{Y.}~\bibnamefont{Zhang}},
  \bibinfo{author}{\bibfnamefont{S.~V.} \bibnamefont{Dubonos}},
  \bibinfo{author}{\bibfnamefont{I.~V.} \bibnamefont{Grigorieva}},
  \bibnamefont{and} \bibinfo{author}{\bibfnamefont{A.~A.}
  \bibnamefont{Firsov}}, \bibinfo{journal}{Science}
  \textbf{\bibinfo{volume}{306}}, \bibinfo{pages}{666} (\bibinfo{year}{2004}).

\bibitem{geimreview}
\bibinfo{author}{\bibfnamefont{A.~K.} \bibnamefont{Geim}} \bibnamefont{and}
  \bibinfo{author}{\bibfnamefont{K.~S.} \bibnamefont{Novoselov}},
  \bibinfo{journal}{Nat. Materials} \textbf{\bibinfo{volume}{6}},
  \bibinfo{pages}{183} (\bibinfo{year}{2007}).

\bibitem{castronetoreview}
\bibinfo{author}{\bibfnamefont{A.~H.} \bibnamefont{Castro~Neto}},
  \bibinfo{author}{\bibfnamefont{F.}~\bibnamefont{Guinea}},
  \bibinfo{author}{\bibfnamefont{N.~M.~R.} \bibnamefont{Peres}},
  \bibinfo{author}{\bibfnamefont{K.~S.} \bibnamefont{Novoselov}},
  \bibnamefont{and} \bibinfo{author}{\bibfnamefont{A.~K.} \bibnamefont{Geim}},
  \bibinfo{journal}{Rev. Mod. Phys.} \textbf{\bibinfo{volume}{81}},
  \bibinfo{pages}{109} (\bibinfo{year}{2009}).

\bibitem{sdsharmareview}
\bibinfo{author}{\bibfnamefont{S.~D.} \bibnamefont{Sarma}},
  \bibinfo{author}{\bibfnamefont{S.}~\bibnamefont{Adam}},
  \bibinfo{author}{\bibfnamefont{E.~H.} \bibnamefont{Hwang}}, \bibnamefont{and}
  \bibinfo{author}{\bibfnamefont{E.}~\bibnamefont{Rossi}},
  \emph{\bibinfo{title}{Electronic transport in two dimensional graphene}}
  (\bibinfo{year}{2010}), \bibinfo{note}{{{arXiv:1003.4731
  [cond-mat.mes-hall]}}}.

\bibitem{katsnelson}
\bibinfo{author}{\bibfnamefont{M.~I.} \bibnamefont{Katsnelson}},
  \bibinfo{author}{\bibfnamefont{K.~S.} \bibnamefont{Novoselov}},
  \bibnamefont{and} \bibinfo{author}{\bibfnamefont{A.~K.} \bibnamefont{Geim}},
  \bibinfo{journal}{Nat. Phys.} \textbf{\bibinfo{volume}{2}},
  \bibinfo{pages}{620} (\bibinfo{year}{2006}).

\bibitem{hubert}
\bibinfo{author}{\bibfnamefont{H.~B.} \bibnamefont{Heersche}},
  \bibinfo{author}{\bibfnamefont{P.~J.} \bibnamefont{Herrero}},
  \bibinfo{author}{\bibfnamefont{J.~B.} \bibnamefont{Oostinga}},
  \bibinfo{author}{\bibfnamefont{L.~M.~K.} \bibnamefont{Vandersypen}},
  \bibnamefont{and} \bibinfo{author}{\bibfnamefont{A.~F.}
  \bibnamefont{Morpurgo}}, \bibinfo{journal}{Nature}
  \textbf{\bibinfo{volume}{446}}, \bibinfo{pages}{56} (\bibinfo{year}{2007}).

\bibitem{andreev}
\bibinfo{author}{\bibfnamefont{A.~F.} \bibnamefont{Andreev}},
  \bibinfo{journal}{Sov. Phys. JETP} \textbf{\bibinfo{volume}{19}},
  \bibinfo{pages}{1228} (\bibinfo{year}{1964}).

\bibitem{beenakker1}
\bibinfo{author}{\bibfnamefont{C.~W.~J.} \bibnamefont{Beenakker}},
  \bibinfo{journal}{Phys. Rev. Lett.} \textbf{\bibinfo{volume}{97}},
  \bibinfo{pages}{067007} (\bibinfo{year}{2006}).

\bibitem{beenakkerreview}
\bibinfo{author}{\bibfnamefont{C.~W.~J.} \bibnamefont{Beenakker}},
  \bibinfo{journal}{Rev. Mod. Phys.} \textbf{\bibinfo{volume}{80}},
  \bibinfo{pages}{1337} (\bibinfo{year}{2008}).

\bibitem{blonder}
\bibinfo{author}{\bibfnamefont{G.~E.} \bibnamefont{Blonder}},
  \bibinfo{author}{\bibfnamefont{M.}~\bibnamefont{Tinkham}}, \bibnamefont{and}
  \bibinfo{author}{\bibfnamefont{T.~M.} \bibnamefont{Klapwijk}},
  \bibinfo{journal}{Phys. Rev. B} \textbf{\bibinfo{volume}{25}},
  \bibinfo{pages}{4515} (\bibinfo{year}{1982}).

\bibitem{subhro}
\bibinfo{author}{\bibfnamefont{S.}~\bibnamefont{Bhattacharjee}}
  \bibnamefont{and} \bibinfo{author}{\bibfnamefont{K.}~\bibnamefont{Sengupta}},
  \bibinfo{journal}{Phys. Rev. Lett.} \textbf{\bibinfo{volume}{97}},
  \bibinfo{pages}{217001} (\bibinfo{year}{2006}).

\bibitem{moitri1}
\bibinfo{author}{\bibfnamefont{S.}~\bibnamefont{Bhattacharjee}},
  \bibinfo{author}{\bibfnamefont{M.}~\bibnamefont{Maiti}}, \bibnamefont{and}
  \bibinfo{author}{\bibfnamefont{K.}~\bibnamefont{Sengupta}},
  \bibinfo{journal}{Phys. Rev. B} \textbf{\bibinfo{volume}{76}},
  \bibinfo{pages}{184514} (\bibinfo{year}{2007}).

\bibitem{linder1}
\bibinfo{author}{\bibfnamefont{J.}~\bibnamefont{Linder}} \bibnamefont{and}
  \bibinfo{author}{\bibfnamefont{A.}~\bibnamefont{Sudb\o{}}},
  \bibinfo{journal}{Phys. Rev. Lett.} \textbf{\bibinfo{volume}{99}},
  \bibinfo{pages}{147001} (\bibinfo{year}{2007}).

\bibitem{linder2}
\bibinfo{author}{\bibfnamefont{J.}~\bibnamefont{Linder}} \bibnamefont{and}
  \bibinfo{author}{\bibfnamefont{A.}~\bibnamefont{Sudb\o{}}},
  \bibinfo{journal}{Phys. Rev. B} \textbf{\bibinfo{volume}{77}},
  \bibinfo{pages}{064507} (\bibinfo{year}{2008}).

\bibitem{beenakker2}
\bibinfo{author}{\bibfnamefont{M.}~\bibnamefont{Titov}} \bibnamefont{and}
  \bibinfo{author}{\bibfnamefont{C.~W.~J.} \bibnamefont{Beenakker}},
  \bibinfo{journal}{Phys. Rev. B} \textbf{\bibinfo{volume}{74}},
  \bibinfo{pages}{041401(R)} (\bibinfo{year}{2006}).

\bibitem{moitri2}
\bibinfo{author}{\bibfnamefont{M.}~\bibnamefont{Maiti}} \bibnamefont{and}
  \bibinfo{author}{\bibfnamefont{K.}~\bibnamefont{Sengupta}},
  \bibinfo{journal}{Phys. Rev. B} \textbf{\bibinfo{volume}{76}},
  \bibinfo{pages}{054513} (\bibinfo{year}{2007}).

\bibitem{greenbaum}
\bibinfo{author}{\bibfnamefont{D.}~\bibnamefont{Greenbaum}},
  \bibinfo{author}{\bibfnamefont{S.}~\bibnamefont{Das}},
  \bibinfo{author}{\bibfnamefont{G.}~\bibnamefont{Schwiete}}, \bibnamefont{and}
  \bibinfo{author}{\bibfnamefont{P.}~\bibnamefont{Silvestrov}},
  \bibinfo{journal}{Phys. Rev. B} \textbf{\bibinfo{volume}{75}},
  \bibinfo{pages}{195437} (\bibinfo{year}{2007}).

\bibitem{cayssol}
\bibinfo{author}{\bibfnamefont{J.}~\bibnamefont{Cayssol}},
  \bibinfo{journal}{Phys. Rev. Lett.} \textbf{\bibinfo{volume}{100}},
  \bibinfo{pages}{147001} (\bibinfo{year}{2008}).

\bibitem{cbenjamin}
\bibinfo{author}{\bibfnamefont{C.}~\bibnamefont{Benjamin}} \bibnamefont{and}
  \bibinfo{author}{\bibfnamefont{J.~K.} \bibnamefont{Pachos}},
  \bibinfo{journal}{Phys. Rev. B} \textbf{\bibinfo{volume}{78}},
  \bibinfo{pages}{235403} (\bibinfo{year}{2008}).

\bibitem{vfalko1}
\bibinfo{author}{\bibfnamefont{V.~V.} \bibnamefont{Cheianov}} \bibnamefont{and}
  \bibinfo{author}{\bibfnamefont{V.~I.} \bibnamefont{Falko}},
  \bibinfo{journal}{Phys. Rev. B} \textbf{\bibinfo{volume}{74}},
  \bibinfo{pages}{041403(R)} (\bibinfo{year}{2006}).

\bibitem{peeters}
\bibinfo{author}{\bibfnamefont{J.~M.} \bibnamefont{Pereira}},
  \bibinfo{author}{\bibfnamefont{P.}~\bibnamefont{Vasilopoulos}},
  \bibnamefont{and} \bibinfo{author}{\bibfnamefont{F.~M.}
  \bibnamefont{Peeters}}, \bibinfo{journal}{Appl. Phys. Lett.}
  \textbf{\bibinfo{volume}{90}}, \bibinfo{pages}{132122}
  (\bibinfo{year}{2007}).

\bibitem{baizhang}
\bibinfo{author}{\bibfnamefont{C.}~\bibnamefont{Bai}},
  \bibinfo{author}{\bibfnamefont{Y.}~\bibnamefont{Yang}}, \bibnamefont{and}
  \bibinfo{author}{\bibfnamefont{X.}~\bibnamefont{Zhang}},
  \bibinfo{journal}{Physica E} \textbf{\bibinfo{volume}{42}},
  \bibinfo{pages}{1431} (\bibinfo{year}{2009}).

\bibitem{beenakker3}
\bibinfo{author}{\bibfnamefont{C.~W.~J.} \bibnamefont{Beenakker}},
  \bibinfo{author}{\bibfnamefont{A.~R.} \bibnamefont{Akhmerov}},
  \bibinfo{author}{\bibfnamefont{P.}~\bibnamefont{Recher}}, \bibnamefont{and}
  \bibinfo{author}{\bibfnamefont{J.}~\bibnamefont{Tworzydlo}},
  \bibinfo{journal}{Phys. Rev. B} \textbf{\bibinfo{volume}{77}},
  \bibinfo{pages}{075409} (\bibinfo{year}{2008}).

\bibitem{morpurgo}
\bibinfo{author}{\bibfnamefont{A.~F.} \bibnamefont{Morpurgo}} \bibnamefont{and}
  \bibinfo{author}{\bibfnamefont{F.}~\bibnamefont{Beltram}},
  \bibinfo{journal}{Phys. Rev. B} \textbf{\bibinfo{volume}{50}},
  \bibinfo{pages}{1325} (\bibinfo{year}{1994}).

\bibitem{arijitkundu}
\bibinfo{author}{\bibfnamefont{A.}~\bibnamefont{Kundu}},
  \bibinfo{author}{\bibfnamefont{S.}~\bibnamefont{Rao}}, \bibnamefont{and}
  \bibinfo{author}{\bibfnamefont{A.}~\bibnamefont{Saha}},
  \bibinfo{journal}{Euro. Phys. Lett.} \textbf{\bibinfo{volume}{88}},
  \bibinfo{pages}{57003} (\bibinfo{year}{2009}).

\bibitem{phillipkim}
\bibinfo{author}{\bibfnamefont{A.~F.} \bibnamefont{Young}} \bibnamefont{and}
  \bibinfo{author}{\bibfnamefont{P.}~\bibnamefont{Kim}}, \bibinfo{journal}{Nat.
  Phys.} \textbf{\bibinfo{volume}{5}}, \bibinfo{pages}{222}
  (\bibinfo{year}{2009}).

\bibitem{cayssol1}
\bibinfo{author}{\bibfnamefont{J.}~\bibnamefont{Cayssol}},
  \bibinfo{author}{\bibfnamefont{B.}~\bibnamefont{Huard}}, \bibnamefont{and}
  \bibinfo{author}{\bibfnamefont{D.}~\bibnamefont{Goldhaber-Gordon}},
  \bibinfo{journal}{Phys. Rev. B} \textbf{\bibinfo{volume}{79}},
  \bibinfo{pages}{075428} (\bibinfo{year}{2009}).

\bibitem{sonin}
\bibinfo{author}{\bibfnamefont{E.~B.} \bibnamefont{Sonin}},
  \bibinfo{journal}{Phys. Rev. B} \textbf{\bibinfo{volume}{79}},
  \bibinfo{pages}{195438} (\bibinfo{year}{2009}).

\end{thebibliography}
\end{document}